\documentclass[twocolumn,prl,showpacs,preprintnumbers,amssymb,superscriptaddress]{revtex4}

\usepackage{epsfig}% Include figure files
\usepackage{dcolumn}% Align table columns on decimal point
\usepackage{bm}% bold math

\usepackage{latexsym}
\usepackage{epsfig,amssymb,euscript,slashed}
\usepackage{array,calc,epsfig}
\usepackage[linktocpage]{hyperref}
\usepackage{pstricks}
\usepackage{bbm}
\usepackage{fancybox}

\def\d {{\rm d}}
\def\be{\begin{equation}}
\def\ee{\end{equation}}

\def\bea{\begin{eqnarray}}
\def\eea{\end{eqnarray}}
\def\bseq{\begin{subequations}}
\def\eseq{\end{subequations}}
\newcommand\bbone{\ensuremath{\mathbbm{1}}}

\newcommand{\nn}{\nonumber}
\newcommand{\beal}{\begin{align}}

\def\calm         {{\cal M}}

\def\calo         {{\cal O}}

\def\cals         {{\cal S}}

\def\calz         {{\cal Z}}

\def\reals        {{\mathbb R}}

\def\tr           {\mathop{\rm Tr}}

\def\sqr#1#2{{\vcenter{\vbox{\hrule height.#2pt
 \hbox{\vrule width.#2pt height#1pt \kern#1pt \vrule width.#2pt}\hrule
 height.#2pt}}}}

\def\oh{\frac{1}{2}}
\def\a{{\alpha}}
\def\b{{\beta}}

\def\eps{{\epsilon}}

\def\lam{{\lambda}}
\def\Om{{\Omega}}

\def\p{{\partial}}
\def\raw{\rightarrow}

\def\slashchar#1{\setbox0=\hbox{$#1$}           % set a box for #1
\dimen0=\wd0                                 % and get its size
\setbox1=\hbox{/} \dimen1=\wd1               % get siste of /
\ifdim\dimen0>\dimen1                        % #1 is bigger
\rlap{\hbox to \dimen0{\hfil/\hfil}}      % so center / in box
#1                                        % and print #1
\else                                        % / is bigger
\rlap{\hbox to \dimen1{\hfil$#1$\hfil}}   % so center #1
/                                         % and print /
\fi}

\begin{document}

\preprint{CERN-PH-TH/2009-195}
\preprint{LMU-ASC 46/09}
%\twocolumn[\hsize\textwidth\columnwidth\hsize\csname@twocolumnfalse\endcsname

\title{Non-perturbative effects on seven-brane Yukawa couplings}

\author{Fernando Marchesano}

\affiliation{CERN PH-TH Division, CH-1211 Geneva, Switzerland}

\author{Luca Martucci}

\affiliation{ASC, LMU-M\"unchen, Theresienstra\ss e 37, D-80333 Munich, Germany}

\begin{abstract}
We analyze non-perturbative corrections to the superpotential of seven-brane gauge theories on type IIB and F-theory warped Calabi-Yau compactifications. We show in particular that such corrections modify the holomorphic Yukawa couplings by an exponentially suppressed contribution, generically solving the Yukawa rank-one problem of certain F-theory local models. We provide explicit expressions for the non-perturbative correction to the seven-brane superpotential, and check that it is related to a non-commutative deformation to the tree-level superpotential via a Seiberg-Witten map.
\end{abstract}


\maketitle

%%%%%%%%%%%%%%%%%%%%%

Within string theory, reproducing the Standard Model of Particle Physics (SM) or extensions thereof has proven to be a complex and challenging quest. This complexity is partly due to the different appearance of string vacua in diverse corners of the string landscape, providing not one, but many possible paths to reproduce the SM as an effective theory. Rather than a drawback, this diversity of scenarios and the web of dualities relating them can be used to render the quest less challenging. Nevertheless, reproducing the qualitative and quantitative features of the SM still remains a non-trivial task.

A good example of the latter is given by the observed hierarchical fermion masses and mixing angles, which any realistic string model should reproduce via an appropriate set of Yukawa couplings. While in each corner of the string landscape the nature and characteristics of Yukawa couplings are quite different, in practice none of the scenarios built so far provides a scheme where a viable set of Yukawas can be derived in a simple, natural way.

In this regard, an interesting arena where such scheme could be developed is the local F-theory scenario recently introduced in \cite{localF}, which realizes the idea of Grand Unification from a bottom-up approach. Indeed, as proposed in \cite{hv08} (see also \cite{if09,cchv09,cp09}), an appealing class of models would be those whose holomorphic Yukawa matrix has rank one, so that just one family of quarks and leptons develops a mass \footnote{The rank-one scenario assumes that down- and up-type Yukawas arise each from a single point of triple intersection of matter curves. Difficulties in building such setting for the latter have been recently pointed out in \cite{obs}.}. While this would be a promising starting point to reproduce the mass hierarchy between the third and first two families of SM fermions, in realistic models the lightest two families need to be massive as well. One then needs to find a source of Yukawas for these two families, which should then provide a small correction to the rank-one piece. While such corrections were initially thought to be built-in within local F-theory constructions, it has been shown in \cite{cchv09} not to be the case, and so in order to solve this problem the tree-level seven brane superpotential $W^{\rm tree}$ of \cite{localF} should be modified by external effects.

The aim of this note is to show that non-perturbative effects can address this rank-one Yukawa problem in the spirit of \cite{hv08} in a rather natural way, by simply adding to $W^{\rm tree}$ a non-perturbatively generated contribution $W^{\rm np}$. In addition, we will argue that non-perturbative effects are the most important source of corrections to the holomorphic Yukawas, at least in the context where the local F-theory models above were initially formulated.

The source of non-perturbative effects modifying the Yukawas will be nothing but the F-theory analogues of type II D-brane instantons, whose effect on 4d effective theories has recently generated a lot of activity \cite{bckw09}. Indeed, that such a mechanism could work was proposed in \cite{ag06}, in the rather different context of the intersecting D6-brane models built in \cite{cim03} and sharing the same rank-one problem. While this initial proposal does not seem to work for such D6-brane models we will see that, when applied to F-theory GUTs, it provides a universal modification $W^{\rm tree} \raw W^{\rm tree} + W^{\rm np}$. Moreover, as the actual expression for $W^{\rm np}$ turns out to be rather simple, this allows to compute its effect in an explicit way, granting the necessary predictive power to the present proposal.

In order to motivate our expression for $W^{\rm np}$ let us first consider type IIB string theory on the warped background $\reals^{1,3} \times_\omega \calm_6$, with O3/O7-planes, as in \cite{gp00,gkp01}. In addition, let us consider $n$ space-time filling D7-branes wrapping a four-cycle $\Sigma_4^{\rm np} \subset \calm_6$, and such that their 4d effective field theory develops a gaugino condensate. If we now consider a D3-brane filling $\reals^{1,3}$ and placed at a point $z_{\rm D3}$ (in complex coordinates) on $\calm_6$ it will develop a superpotential of the form \cite{ganor96,bhk04,bdkmmm06}
\be
W^{\rm np}_{\rm D3}\, =\, \mu_3 {\cal A}\, e^{-T_\Sigma/n} f(z_{\rm D3})^{1/n} 
\label{WnpD3}
\ee
where $T_\Sigma = V_{\Sigma_4^{\rm np}} + i \int_{\Sigma_4^{\rm np}} C_4$ is the complexified K\"ahler modulus corresponding to the four-cycle $\Sigma_4^{\rm np}$, and $f$ stands for a holomorphic section of the divisor bundle specifying $\Sigma_4^{\rm np}$, with no poles and such that it vanishes on $\Sigma_4^{\rm np}$. In addition,  $\mu_p = (2\pi)^{-p} \a'^{-\frac{p+1}{2}}$ is the tension of a D$p$-brane and ${\cal A}$ is a holomorphic function of the complex structure moduli of $\calm_6$. This function can actually be considered a constant if supersymmetric three-form fluxes $G_3 = F_3 + \tau H_3$ are introduced, since they lift the complex structure moduli of $\calm_6$ via a closed string superpotential \cite{gvw99,drs99}. Note, however, that no extra superpotential is generated for the D3-brane in the presence of these fluxes, and so the result (\ref{WnpD3}) remains unaffected. Alternatively, instead of a gaugino condensing D7-brane we may consider an isolated Euclidean D3-brane instanton wrapping the same four-cycle $\Sigma_4^{\rm np}$, provided that it contains the appropriate number of fermionic zero modes \cite{witten96}. The superpotential generated by  such instanton is again given by eq.(\ref{WnpD3}), now with $n=1$. For simplicity, in the following we will focus on this latter possibility.

In order to understand the superpotential (\ref{WnpD3}) in terms of the 4d D3-brane gauge theory, we need to Taylor expand $f(z)$ around the D3-brane location $z_{\rm D3}$, and then express such expansion as D3-brane complex fields $\phi_{\rm D3}^i  = \lam (z^i - z_{\rm D3})$, with $\lam= 2\pi\a'$. We thus obtain 
\be
W^{\rm np}_{\rm D3}\, =\, \mu_3 {\cal A}\,e^{-T_\Sigma}  (f|_{z_{\rm D3}} + \lam \phi^i[\p_{z^i}f]_{z_{\rm D3}} + \dots)
\label{WnpD3exp}
\ee
The same philosophy applies to a stack of $N$ D3-branes. The superpotential (\ref{WnpD3}) naturally generalizes to
\be
W^{\rm np}_{\rm D3} = \mu_3{\cal A}\, e^{-T_\Sigma}[\det f(Z_{\rm D3})] 
\label{WnpND3}
\ee
where $Z_{\rm D3}$ is now made up of $N\times N$ complex matrices. Expanding $Z^i_{\rm D3} = z^i_{\rm D3} \bbone_N+\lam \phi^i$, we have
\be
W^{\rm np}_{\rm D3} = \mu_3{\cal A} e^{-T_\Sigma}\hspace*{-.05cm} f|^N_{z_{\rm D3}}  \{1 + \lam \tr( \phi^i)[\p_{z^i} \log f]_{z_{\rm D3}} + ...\}
\label{WnpND3exp}
\ee
where $N$ is the D3-brane charge of the system.

Let us now replace the D3-branes at $z_{\rm D3}$ by a stack of D7-branes on $\reals^{1,3} \times \cals_4$, where $\cals_4$ is a compact, complex four-cycle with no intersection with $\Sigma_4^{\rm np}$, and with a non-trivial worldvolume flux $F= \d A - \frac{i}2[A,A]$ along $\cals_4$. Since $\cals_4$ and $\Sigma_4^{\rm np}$ are physically separated, we can treat the four-cycle $\cals_4$ as a smeared source of D3-brane, whose total charge is given by $N_{\rm D3} = \int_{\cals_4} \tr (F\wedge F)/8\pi^2 \in \mathbb{N}$. Such smeared source will backreact on both the Ramond-Ramond (RR) potential $C_4$ and the warp factor, implying that the D3-instanton action (whose real part is the warped volume of $\Sigma_4^{\rm np}$) will depend on the D7-brane moduli. Adapting the analysis of \cite{bdkmmm06}, one is led to the conclusion that a D7-brane on $\cals_4$ should develop a non-perturbative superpotential of the form 
\be\label{WnpD7}
W_{\rm D7}^{\rm np}\, =\, \mu_3{\cal A}\,  e^{-T_{\Sigma}} \exp\left[\frac{1}{8\pi^2}{\int_{\cals_4}  {\rm Str}(\log f\, F\wedge F) }\right]
\ee
where Str indicates the symmetric trace. $W_{\rm D7}^{\rm np}$ is  to be added to the tree-level superpotential \cite{luca06}
\bea\label{treeD7}
W^{\rm tree}_{\rm D7} =2\pi\alpha^\prime\mu_7 \int_{\Gamma_5}{\rm Str}\, \Omega\wedge F
\eea
where $\Gamma_5$ is a 5-chain connecting $\cals_4$ and a reference four-cycle, and $F$ is a proper extension of the D7-brane world-volume flux on $\Gamma_5$.

The non-Abelian expressions (\ref{WnpD7}) and (\ref{treeD7}) can be made more precise by expanding them around a holomorphic embedding $\cals_4$. Introducing local coordinates  $(u,v,w)$ such that $\cals_4$ is described by $w=0$, we can expand $W_{\rm D7}^{\rm np}$ in the complex, non-Abelian field $\phi=2\pi \alpha^\prime w$. Since $f|_{\cals_4}$ is a holomorphic function with no poles and zeros on the compact divisor $\cals_4$, it is in fact a complex constant that can be pulled-out of the integral, and so the first term of this expansion will be the constant. Then, up to second order terms in $\phi$, the non-perturbative superpotential (\ref{WnpD7})  is given by
\be
W_{\rm D7}^{\rm np} =\mu_3 {\cal A} e^{-T_\Sigma} f|_{\cals_4}^{N_{\rm D3}} + \frac{\mu_3}{8\pi^2}\int_{\cals_4} \theta\,{\rm Str} (\phi\, F\wedge F) + \dots
\label{WnpD7exp}
\ee
with $\theta:=\lam {\cal A} e^{-T_{\Sigma}}(f^{N_{\rm D3}} \partial_w\log f)|_{\cals_4}$.  Adding up $W^{\rm tree}_{\rm D7}$ and $W^{\rm np}_{\rm D7}$ and neglecting constant contributions we obtain
\be\nn
W_{\rm D7} =\frac{\mu_3}{4\pi^2} \hspace*{-.15cm}  \int_{\cals_4}\hspace*{-.2cm} \big[(\iota_w\Omega) \wedge \tr(\phi\, F)  + \frac12\, \theta\, {\rm Str}(\phi\, F^2)+ \dots\big]
\label{WtotD7} 
\ee
where we have kept only linear terms in $\phi$ \footnote{Taking global properties of the holomorphic  normal bundle  into account, $W^{\rm tree}_{\rm D7}$ may also contain quadratic terms on $\phi$ whenever $F\neq 0$. These terms, not visible in our local analysis, can be easily incorporated into eq.(\ref{WtotD7}).}.  

Besides warping and $C_4$, a D7-brane also sources the dilaton and the RR potential $C_0$. Its backreaction is then more involved than $N_{\rm D3}$ smeared D3-branes, a fact which in principle could complicate the above analysis. However, such extra fields do not enter into the action of an instantonic D3-brane wrapping $\Sigma_4^{\rm np}$, whenever its worldvolume flux $F$ vanishes. Hence, this additional backreaction does not change the computation above, and so the non-perturbative superpotential indeed reduces to (\ref{WnpD7}). The same statement applies to $n$ condensing D7-branes with vanishing worldvolume flux, for which (\ref{WnpD7}) can be trivially extended.

As advanced, the D7-brane superpotential (\ref{WtotD7}) splits as $W_{\rm D7}^{\rm tree} + W_{\rm D7}^{\rm np}$, the first piece being the superpotential considered in \cite{localF} and the second piece a non-perturbative correction. Note that $W^{\rm np}_{\rm D7}$, compared to $W^{\rm tree}_{\rm D7}$, contains an extra factor $\theta$ of dimension (length)$^2$. This compensates the higher dimensional integrand Str$(\phi\, F\wedge F)$ from which we can extract a coupling of up to five fields. Corrections to the tree-level Yukawa couplings then arise from terms involving only three fluctuations, like
\be
\int_{\cals_4} \theta\,\eps^{i\bar{\jmath}k\bar{l}}  {\rm Str}\left(\phi\, D_i A_{\bar{\jmath}} D_k A_{\bar{l}} \right)
\label{correction}
\ee
($D_k = \p_k + i \langle A_k\rangle\wedge$ not containing any fluctuation), as well as from deformations of the tree-level wavefunction profile induced by the presence of $W_{\rm D7}^{\rm np}$.

The corrected superpotential (\ref{WtotD7}) admits an interesting interpretation, inspired by some observations made in \cite{km07}. There, it was proposed to encode non-perturbative corrections to D-brane superpotentials in terms of deformations of the bulk geometry as seen by D-branes. In the case at hand, the correction would be encoded in a $\b$-deformation of the internal complex structure. Indeed, in the type IIB/F-theory backgrounds of \cite{gp00,gkp01} there is an integrable complex structure specified by the holomorphic $(3,0)$-form $\Omega$. A $\beta$-deformation replaces $\Omega$ by the more general pure spinor odd polyform $\calz$ \cite{beta} \footnote{This description of $\beta$-deformations should be understood as a deformation of the type IIB supergravity background in which IASD 3-form fluxes are turned on, as in \cite{gp00}.}
\be
\calz\, =\, \calz_{1} + \calz_{3} \quad \quad {\rm with} \quad \quad \calz_{3} \, =\, \Om, \quad 
\calz_{1}\, =\, \b\lrcorner \Om
\label{polyZ}
\ee
where $(\b\lrcorner \Om)_k =\frac12 \b^{ij}\Om_{ijk}$. Here $\b = \oh \b^{ij} \p_{z^i} \wedge \p_{z^j} + \text{c.c.}$ is a (2,0)+(0,2) real bivector whose $(2,0)$ component is holomorphic, and defines a Poisson structure. Since integrability imposes that $\d\calz=0$, we can locally write  $\calz_3\equiv \Omega = \partial\chi_{2}$ and $\calz_1 = \partial\chi_0$. Hence, using the superpotential for Abelian D7-branes derived in \cite{luca06} we obtain
\be
\label{WlucaD7}
W_{D7}  =  2\pi \a' \mu_7\int_{\cals_4}\hspace*{-.1cm} \left( \pi \a' \chi_{0} F\wedge F+\chi_{2} \wedge F \right)+\text{const.}
\ee
which, appropriately choosing the additional constant, reproduces the Abelian version of (\ref{WtotD7}) by simply taking $\chi_0 = {\cal A} e^{-T_\Sigma} f^{N_{\rm D3}}/N_{\rm D3}$ \footnote{Note that this description breaks down at $\Sigma_4^{\rm np}$, since $\partial\bar\partial \ln|\chi_0|^2$  is a $\delta$-function two-form with support on $\Sigma_4^{\rm np}$.}. Now, as argued in \cite{kapustin03,pestun}, on a D-brane world-volume the effect of a $\beta$-deformation can be seen as a non-commutative deformation of the gauge theory. All this suggests that the non-perturbative correction to the D7-brane superpotential (\ref{WnpD7}) should be equivalent to non-commutative deformation of the tree-level piece, via the standard Seiberg-Witten map \cite{sw99}.

Indeed, to connect our results with those in \cite{kapustin03}, let us consider the superpotential (\ref{WlucaD7}). Recalling that $\d\chi_{2} = \Om$ and $\d\chi_{0} = \b\lrcorner \Om$,  the F-flatness conditions read
\bea
\label{Fflat}
&&\eta|_{\cals_4}\wedge F+\Omega|_{\cals_4} = 0 \\ \nonumber
&&(\iota_X\eta)|_{\cals_4} \, F^2 + 2(\iota_X\Omega)|_{\cals_4}\wedge F = 0  \quad \forall X \in TM|_{\cals_4}
\eea
where $\eta = 2\pi\a' \b\lrcorner \Om$. The first F-flatness condition is automatically satisfied in our previous setup, since $\cals_4$ was chosen to be a divisor (so that $\Om|_{\cals_4} =0$) while $\chi_0$ was a constant function on $\cals_4$, and thus $\eta|_{\cals_4}=0$. The second condition in (\ref{Fflat}) can be rewritten in a more explicit way by taking again the local system of  coordinates $(u, v, w)$. In this system $\b\lrcorner \Om|_{\cals_4} = 0$ implies $\b^{vw}|_{\cals_4} = \b^{uw}|_{\cals_4} = 0$. Then, defining the bivector $\Theta \equiv 2\pi \a'\b|_{\cals_4}$, one can show that the second condition in (\ref{Fflat}) is equivalent to
\be
FI+I^TF=-F(I\Theta+\Theta I^T)F
\label{Fflatk}
\ee
with $I$ the complex structure associated to $\Omega$. As shown in \cite{kapustin03}, eq.(\ref{Fflatk}) is nothing but the current-matching condition for a B-brane in the $\b$-deformed topological theory.

It was shown in \cite{kapustin03} that (\ref{Fflatk}) is equivalent to $\hat{F}^{(0,2)} = 0$, $\hat{F}$ the non-commutative field-strength constructed via the Seiberg-Witten (SW) map \cite{sw99}. We now show that this relation can be extended off-shell, deriving from (\ref{WtotD7}) the non-commutative superpotential used in \cite{cchv09}.\footnote{This derivation not only applies to the non-perturbative superpotential (\ref{WtotD7}), but to any superpotential of the form (\ref{WlucaD7}) arising from general $\beta$-deformed complex spaces.}

Choosing again local coordinates such that $\Om = \d u \wedge \d v \wedge \d w$, we have that $\Theta=\theta\,\partial_u\wedge \partial_v+\text{c.c.}$ 
We would then expect to arrive to a non-commutative superpotential of the form (omitting overall dimensionful constant factors)
 \be
 \label{nancd7}
 \hat W_{\rm D7}= \int_{\cals_4} \tr(\hat\varphi\circledast \hat F )
 \ee
where $\hat\varphi=\hat\phi\,\d u\wedge \d v$,  and $\circledast$ and $\hat F$ are non-commutative deformations of the ordinary wedge-product and field-strength respectively, see below.

Let us start by assuming that we have a constant $\theta=\theta_0$, as in \cite{kapustin03}, so that the standard SW map of \cite{sw99} can be applied.  
In this case $\circledast$ can be simply obtained from the ordinary wedge-product by multiplying the components of forms using the  the ordinary Moyal $*$-product defined by the bivector $\Theta$, and $\hat F_{\alpha\beta}=\partial_\alpha\hat A_\beta-\partial_\beta \hat A_\alpha-i(\hat A_\alpha*\hat A_\beta-\hat A_\beta*\hat A_\alpha)$. We can now apply the non-Abelian SW map
\be\label{SWmap}
\begin{array}{rcl}
 \vspace*{.05cm}
 \hat F_{\alpha\beta}&=&F_{\alpha\beta}+ \Theta^{\gamma\delta}[\{F_{\alpha\gamma}, F_{\delta\beta}\} +\\  \vspace*{.05cm} 
 && \frac12\{A_\gamma,  (D_\delta+\p_\delta) F_{\alpha\beta}\} ] +\calo(\theta^2)
 \\
\hat\phi&=&\phi+\frac12\Theta^{\alpha\beta}\{ A_\alpha, (D_\beta+\p_\beta)\phi)\} +\calo(\theta^2)
\end{array}
\ee
with $D_\alpha F_{\beta\gamma}=\p_\alpha F_{\beta\gamma}-i[A_{\alpha},F_{\beta\gamma}]$, $D_\alpha\phi=\p_\alpha\phi-i[A_\alpha,\phi]$, and where the action of the SW map on scalars $\phi$ can be obtained by consistency with T-duality. Plugging these definitions into (\ref{nancd7}), and keeping only terms up to order $\theta_0$ and $\bar\theta_0$, we indeed get the superpotential (\ref{WtotD7}), providing the equivalence up to this order.

 In order to allow for a non-constant $\theta$, one can follow the strategy of \cite{cchv09},  and choose a holomorphic frame $e_{I}=\{e_U,e_V\}$ (with $[e_I,e_J]=[e_I,e_{\bar J}]=0$) in which $\Theta=\theta_0\,e_U\wedge e_V+\text{c.c.}$, with $\theta_0$ again constant  (see Appendix B of \cite{cchv09} for further details). Then the extension of the ordinary  Moyal-product is given by
\bea\label{moyal}
f*g=f\, e^{\frac{i\theta_0}{2}\epsilon^{IJ}(\overleftarrow{e}_I\otimes\overrightarrow{e}_J)}e^{\frac{i\bar \theta_0}{2}\epsilon^{\bar I\bar J}(\overleftarrow{ e}_{\bar I}\otimes\overrightarrow{e}_{\bar J})}\, g
\eea 
and the non-commutative wedge-product $\circledast$  by expanding the forms in the coframe $e^{I},e^{\bar I}$ and applying the above $*$-product to the components. For instance, given a (1,0)-form $\alpha=\alpha_I e^I$ and a (0,1)-form $\beta=\beta_{\bar I}e^{\bar I}$ we have
\bea
\alpha\circledast\beta=(\alpha_{I}*\beta_{\bar J})\, e^{I}\wedge e^{\bar J}
\eea
Working in the basis $e_{I},e_{\bar I}$, one can thus extend the above SW map to these cases with non-constant $\theta$ and show that, up to first-order in $\theta$, the non-commutative superpotential (\ref{nancd7}) is equivalent to (\ref{WtotD7}). 

Note that the above non-commutative products $*$ and $\circledast$ (when applied to non-holomorphic functions) do not coincide with the ones introduced in \cite{cchv09}, denoted $*_{\rm h}$ and $\circledast_{\rm h}$ in the following. The key difference is that $*_{\rm h}$ and $\circledast_{\rm h}$ involve only the holomorphic $(2,0)$ component of $\Theta$, $\Theta^{2,0}=\theta\partial_u\wedge \partial_v$. For instance,
 \bea\label{moyalhol}
f*_{\rm h}g=f\, e^{\frac{i\theta_0}{2}\epsilon^{IJ}(\overleftarrow{e}_I\otimes\overrightarrow{e}_J)}\, g
\eea 
In particular, \cite{cchv09} used the following superpotential
 \be
 \label{cecsup}
 \tilde W_{\rm D7}= \int_{\cals_4} \tr(\tilde\varphi\circledast_{\rm h} \tilde F )
 \ee
with $\tilde F^{0,2}=\bar\partial A^{0,1}-\frac{i}2[A^{0,1},A^{0,1}]_{*_{\rm h}}$,
in order to solve the rank-one Yukawa problem. It would thus seem that both deformations of the tree-level superpotential are unrelated. This is however not the case, since (\ref{nancd7}) and (\ref{cecsup}) are related by an anti-holomorphic SW map.  Let us explicitly discuss the case of constant $\bar\theta=\bar\theta_0$, the general case being analogous by the remarks of the previous paragraph. Then the anti-holomorphic SW map, taking $\tilde F^{0,2}$ and $\tilde\phi$ into $\hat F^{0,2}$ and $\hat\phi$, is again of the form (\ref{SWmap}), with the substitutions $\Theta \raw \Theta^{0,2}=\bar\theta_0\partial_{\bar u}\wedge \partial_{\bar v}$, $F \raw \tilde F^{0,2}$, $\phi\raw \tilde \phi$ and all products built with $*_{\rm h}$ on its rhs. This map indeed preserves the complexified gauge transformations $\hat \delta \hat A^{0,1}={\bar\partial}\hat\lambda-i[\hat A^{0,1},\hat\lambda]_{\circledast}$ and $\tilde \delta \tilde A^{0,1}={\bar\partial}\tilde\lambda-i[\tilde A^{0,1},\tilde\lambda]_{\circledast_{\rm h}}$, which are symmetries of  (\ref{nancd7})  and  (\ref{cecsup}) respectively. One can then check that, up to first order in $\bar\theta_0$  and to all orders in $\theta_0$, (\ref{nancd7})  is indeed mapped to (\ref{cecsup}).

To summarize, we have provided evidence that non-perturbative effects generated by Euclidean D3-branes or gaugino condensing D7-branes produce a simple but interesting correction to the superpotential of D7-brane gauge theories. Moreover, by applying the approach of \cite{bdkmmm06} to magnetized D7-branes, we have derived an explicit, general expression for such corrections. It would however be interesting to check this result by means of a direct CFT computation, along the lines of \cite{ag06,bhk04}.

Even if our discussion was carried in the type IIB context, it can be easily extended to F-theory. In particular, it can be applied to F-theory GUT models with rank-one Yukawas in order to lift their degeneracies. Note that non-perturbative corrections to the tree-level Yukawa matrix are exponentially suppressed, so they can be treated as a small correction to $Y_{ijk}^{\rm tree}$. Moreover, the simple expression obtained for $W^{\rm np}$ allows to carry a systematic analysis of the textures that $W^{\rm np}$ may give rise to, a task that we leave for future work.

We have also identified via a SW map this non-pertur\-bative correction with a non-commutative deformation of the initial, tree-level superpotential. We have in particular recovered, to first order in $\theta$, the non-commutative deformation considered in \cite{cchv09}. As pointed out there, such deformation  generically solves the Yukawa rank-one problem in F-theory, in agreement with our expectations. In \cite{cchv09} the source for such non-commutative deformation was advocated to a tree-level effect due to the presence of background 3-form fluxes. This possibility is however excluded for the no-scale F-theory flux backgrounds of \cite{bb96,gkp01} where the models of \cite{localF,cchv09} are formulated. Indeed, we have seen that the non-commutative deformation (\ref{cecsup}) is equivalent to a seven-brane superpotential piece of the form (\ref{WnpD7exp}), and it is easy to convince oneself that \cite{luca06}
\be\label{explicit}
W_{\rm D7}^{\rm tree} \supset \int_{\cals_4} {\rm Str} (\chi_0 F\wedge F) \iff  W_{\rm D3}^{\rm tree} = \chi_0
\ee
$\chi_0$ being a holomorphic function to be Taylor expanded. %to obtain (\ref{WnpD7exp}). 
In \cite{bb96,gkp01}, $W_{\rm D3}^{\rm tree} = 0$ and D3-brane superpotentials can only be generated at the non-perturbative level \cite{bdkmmm06}, so such non-commutative deformation can only have a non-perturbative origin, as we obtain from our setting.

\acknowledgments

%\vspace*{.2cm}

\smallskip

\centerline{\small \bf Acknowledgements}

%\vspace*{.1cm}

\smallskip 

\vspace*{-.1cm}

We wish to thank M.~Haack, L.~Ib\'a\~nez, M.~Schmidt-Sommerfeld and A.~Uranga for useful discussions.
L.M.'s work is supported  by the DFG Cluster of Excellence ``Origin and Structure of  the Universe'' in Munich, Germany.

\end{document}